\documentclass[12pt, a4paper]{scrartcl}

\usepackage[utf8]{inputenc}
\usepackage[T1]{fontenc}
\usepackage[USenglish]{babel}

\usepackage[tracking=true, protrusion=true, expansion=true,final, babel]{microtype}
\usepackage{lmodern}

\usepackage[pdfborderstyle={/S/U/W 1}]{hyperref}

\usepackage[version=4]{mhchem} 

\usepackage{amssymb}

\usepackage{graphicx}
\usepackage[labelfont=bf]{caption}
\usepackage{placeins}

\usepackage[separate-uncertainty]{siunitx}
\DeclareSIUnit\at{at.\%}
\DeclareSIUnit\rpm{rpm}

\usepackage[style=numeric, maxbibnames=99, sorting=none, citestyle=numeric-comp, giveninits=true]{biblatex}
\usepackage{csquotes}

\usepackage{authblk}

\title{Near-Atomic Scale Perspective on the Oxidation of \ce{Ti3C2T_x} MXenes: Insights from Atom Probe Tomography}
\date{}

\author[1]{Mathias Krämer}
\author[2]{Bar Favelukis}
\author[1,3]{Ayman A. El-Zoka}
\author[2]{Maxim Sokol}
\author[2]{Brian A. Rosen}
\author[2]{Noam Eliaz}
\author[1,4,*]{Se-Ho Kim}
\author[1,3,*]{Baptiste Gault}

\affil[1]{Max-Planck-Institut für Eisenforschung, Max-Planck-Straße 1, 40237 Düsseldorf, Germany}

\affil[2]{Department of Materials Science and Engineering, Tel Aviv University, P.O.B 39040, Ramat Aviv 6997801, Israel}

\affil[3]{Department of Materials, Royal School of Mines, Imperial College London, London, SW7 2AZ, United Kingdom}

\affil[4]{Department of Materials Science and Engineering, Korea University, Seoul 02841, Republic of Korea}

\affil[*]{Corresponding authors: sehonetkr@korea.ac.kr, b.gault@mpie.de}

\begin{document}

\maketitle


\clearpage

\section*{Abstract}

MXenes are a family of 2D transition metal carbides and nitrides with remarkable properties and great potential for energy storage and catalysis applications. However, their oxidation behavior is not yet fully understood, and there are still open questions regarding the spatial distribution and precise quantification of surface terminations, intercalated ions, and possible uncontrolled impurities incorporated during synthesis and processing. Here, atom probe tomography analysis of as-synthesized \ce{Ti3C2T_x} MXenes reveals the presence of alkali (\ce{Li}, \ce{Na}) and halogen (\ce{Cl}, \ce{F}) elements as well as unetched \ce{Al}. Following oxidation of the colloidal solution of MXenes, it is observed that the alkalies enriched in \ce{TiO2} nanowires. Although these elements are tolerated through the incorporation by wet chemical synthesis, they are often overlooked when the activity of these materials is considered, particularly during catalytic testing. This work demonstrates how the capability of atom probe tomography to image these elements in 3D at the near-atomic scale can help to better understand the activity and degradation of MXenes, in order to guide their synthesis for superior functional properties. 

\section*{Keywords}

atom probe tomography, 2D materials, MXene, nanowires, oxidation


\clearpage

\section*{}

The discovery of MXenes by the Barsoum and Gogotsi groups in 2011 has reignited scientific interest in 2D and layered materials~\cite{2011Nag}. To date, numerous MXenes have been synthesized~\cite{2021Nag}, and are being intensively studied and tested in applications including supercapacitors and batteries~\cite{2017Ana, 2022Li}, optoelectronics~\cite{2021Mal}, catalysis~\cite{2019LiZhe}, and biosensors or antibacterial membranes~\cite{2018Hua_CSR, 2019Sol}. Their properties and performance in various applications are significantly affected by unavoidable surface terminations, denoted by \ce{T_x} in the chemical formula, which saturate the bare MXene surface during synthesis~\cite{2022Lim}. Changes in the electronic properties of the MXenes, for example, have been explained by careful consideration of these functional groups and possible intercalants originating from wet chemical synthesis~\cite{2019Har}. Using them to fine-tune properties would require establishing relationships between activity and their spatial distribution and concentration, which remain extremely difficult to characterize.

Atom Probe Tomography (APT) is one of the few techniques that can measure composition in 3D at the near-atomic scale~\cite{2021Gau}. In recent years, the application of APT to nanoparticles~\cite{2011Ted, 2014Li, 2014Fel}, nanowires~\cite{2006Per, 2020Lim} and nanosheets~\cite{2020Kim} has triggered a reckoning that the detailed composition and the presence of trace impurities are of paramount importance when it comes to understanding the activity of nanomaterials~\cite{2022Li_Cell}. Impurity elements in the raw materials used during the synthesis may not react at the same rate and hence be preferentially incorporated into the final product~\cite{2020Kim}, or may be incorporated into the structure during processing or synthesis by wet chemistry~\cite{2022Kim_JACS, 2022Kim_AM}. For MXenes and their surface terminations, most studies to date have relied on X-ray spectroscopy and electron microscopy techniques for compositional measurements~\cite{2021Nat, 2021Aln, 2019Nat, 2020Lee, 2022Ngu}, leaving many questions unanswered that could be potentially be addressed by using APT. However, APT has never been applied to MXenes before.  

As the structural and chemical stability of MXenes can also be extremely fragile, APT could also be a tool for better understanding the mechanisms behind the degradation. MXenes degrade particularly fast in an oxidizing environment (\ce{H2O}, \ce{O2}, etc.), leading to their transformation into transition metal oxides~\cite{2021Iqb, 2022Jia}. Oxidation, in its early stage, can affect the surface terminations~\cite{2020Per} and thus the properties. For example, an increased concentration of \ce{O} functional groups results in an increased catalytic activity~\cite{2017Gao}. Several studies have demonstrated how further oxidation severely destroys the integrity of the MXene structure, thereby limiting its lifetime in service and precluding many useful applications due to a decrease in the properties of interest~\cite{2022Cao, 2019LiXin, 2020Lia, 2021Bha}. Although efforts have been made to understand the details of their evolving chemistry during oxidation~\cite{2014Gha, 2022Bad}, some questions are still unanswered, such as the role of non-\ce{O} surface terminations, intercalants or impurities from synthesis. Designing MXenes with improved oxidation resistance is therefore extremely challenging, and for this reason the improvement of chemical and temperature stability of MXenes is considered in the community to be one of the research challenges of this decade~\cite{2021Gog}.

In order to further the understanding of the chemical evolution of colloidal \ce{Ti3C2T_x} MXene solution, we introduce here the first APT analyses of as-synthesized MXenes, alongside with the transition metal oxides formed during oxidation. APT revealed the incorporation of alkali (\ce{Li}, \ce{Na}) and halogen (\ce{Cl}, \ce{F}) elements into the MXene nanosheets, probably inherited from the synthesis. Following oxidation, these elements remained within the newly formed \ce{TiO2} nanowires, indicating that they lowered their free energy and helped to stabilize them. Although most of these elements are considered to be an essential part of the MXenes themselves, these results emphasize that discussions of the activity of MXene-based materials, including throughout oxidation, should take account of them, which are likely to play an important role in the activity~\cite{2021Wes} and their resistance to degradation.

\begin{figure}[!htb]
    \centering
    \includegraphics[width = 7.01 in]{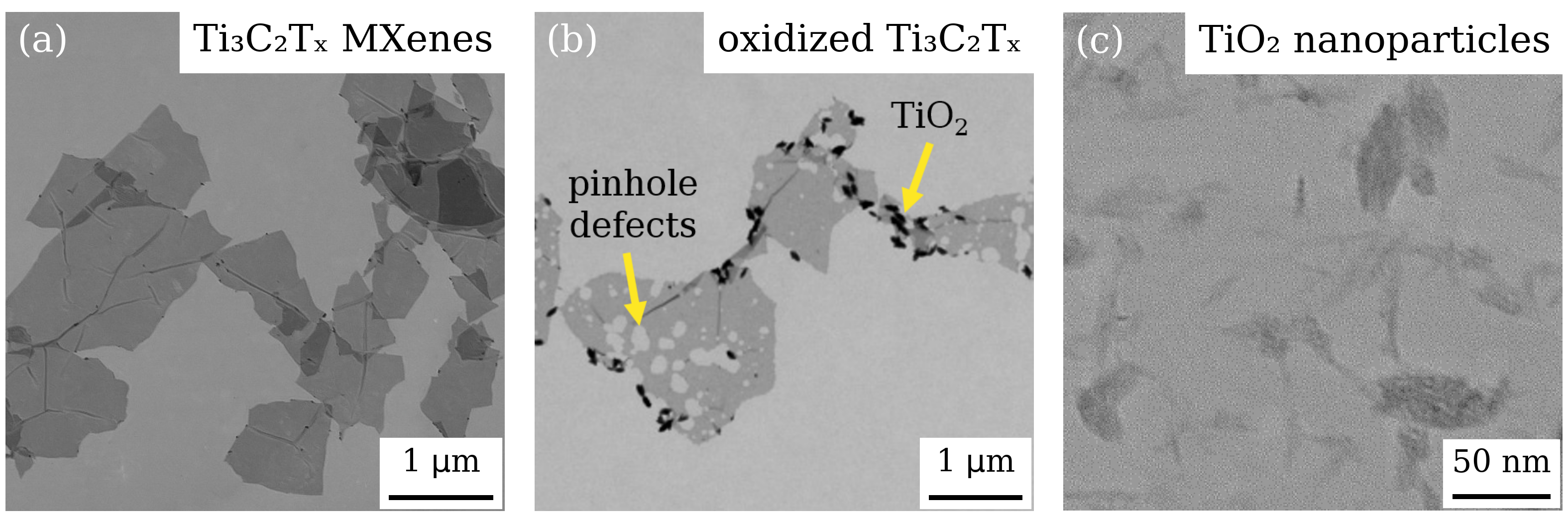}
    \caption{Oxidation processes of the \ce{Ti3C2T_x} MXenes. (a) As-synthesized \ce{Ti3C2T_x} MXene nanosheets. (b) Oxidized \ce{Ti3C2T_x} MXene after storage in a colloidal solution in a refrigerator. (c) \ce{TiO2} nanoparticles as products of \ce{Ti3C2T_x} MXene oxidation.}
    \label{fig:Synthesis_Oxidation}
\end{figure}

\ce{Ti3C2T_x} MXenes were synthesized by wet chemical etching of the \ce{Al} layer from a \ce{Ti3AlC2} MAX phase via the \ce{HCl}-\ce{LiF} route (synthesis of both materials is described in detail in the Supporting Information). In \autoref{fig:Synthesis_Oxidation}~(a), the as-synthesized MXenes shows very clean edges and a defect-free surface, suggesting no major internal defects. Small black dots on the edges may indicate the formation of oxidation products immediately after synthesis. To observe the oxidation of the nanosheets with electron microscopy, the colloidal \ce{Ti3C2T_x} MXene solution was kept at a temperature below \SI{5}{\degreeCelsius} throughout the oxidation experiment, as described in Ref.~\cite{2019Cha}.

Oxidation of the nanosheets in the colloidal solution may have started from the edges due to local structural variations~\cite{2015Kar}, where nanoparticles were observed in \autoref{fig:Synthesis_Oxidation}~(b). Previous research has reported the formation of carbon-supported anatase during the oxidation of \ce{Ti3C2T_x} MXenes~\cite{2014Nag}, indicating that the nanoparticles formed are \ce{TiO2}. Because atomic vacancies and microstructural defects such as wrinkles are unavoidable in the course of the wet chemical synthesis processes, these preexisting defects on the \ce{Ti3C2T_x} MXene may have created a local electric field that drives the development of the \ce{TiO2} particles by promoting the migration of both \ce{Ti} cations and electrons~\cite{2019Fan}. These pinhole defects then gradually evolved into larger voids, resulting in the degradation and accelerated oxidation of the MXene structure. Zhang et al.~\cite{2017Zha} reported that oxidation starts at the defective edges of the MXene sheets and propagates inwards, causing cracks to nucleate and grow. This so-called 'scissor effect' shreds the MXene sheets into small pieces and leads to a complete loss of the original structure of the MXene into \ce{TiO2} debris, such as those imaged in \autoref{fig:Synthesis_Oxidation}~(c). 

Dense and sharp needle-shaped APT specimens containing either the as-synthesized or the oxidized \ce{Ti3C2T_x} MXenes were prepared from a nanosheet-metal composite (detailed APT specimen preparation is provided in the Supporting Information). The electrodeposition within a metallic matrix, used as an encapsulating material, increases the success rate and data quality~\cite{2018Kim, 2021Jun}. Here, a \ce{Co} matrix was used, which, compared to the more commonly used \ce{Ni}, presents the advantage of having only a single isotope, making it less likely to create peaks that can obscure the signal from the material of interest in the APT mass spectrum. For example, the use of \ce{Ni} would cause an unavoidable peak overlap with \ce{TiO} molecular ions in the APT mass spectrum~\cite{2020Lim}.

\begin{figure}[!htb]
    \centering
    \includegraphics[width = 7.01 in]{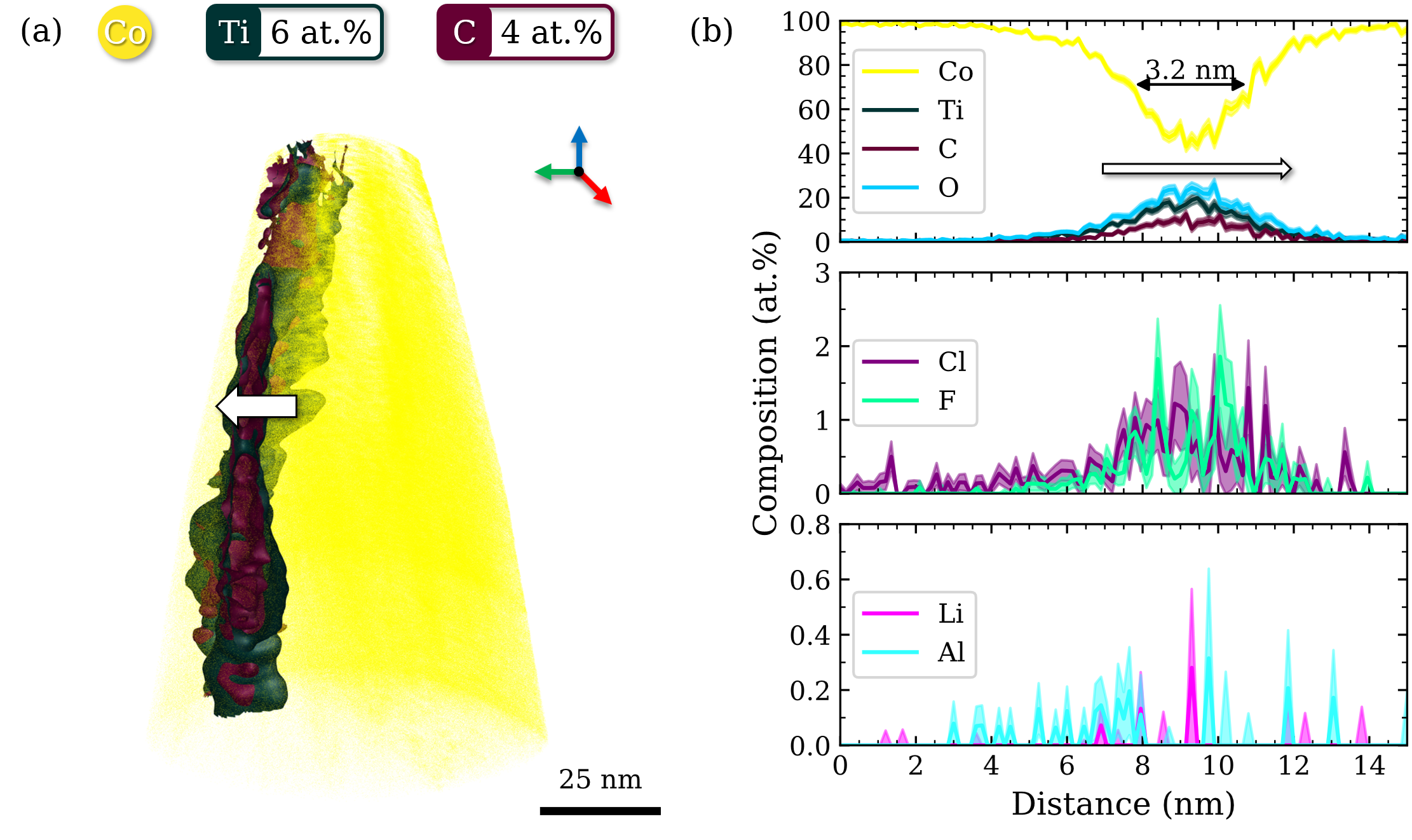}
    \caption{APT analysis of as-synthesized \ce{Ti3C2T_x} MXenes. (a) Reconstructed 3D atom map. Agglomerated \ce{Ti3C2T_x} MXene nanosheets are highlighted by a dark green-cyan iso-compositional surface at \SI{6}{\at} \ce{Ti} and a reddish-purple iso-compositional surface at \SI{4}{\at} \ce{C}. (b) 1D compositional profile ($\varnothing$\SI{15}{\nm}~x~\SI{15}{\nm}) across the agglomerated MXenes as indicated in (a).}
    \label{fig:APT_analysis_Ti3C2}
\end{figure}

The reconstructed 3D atom map in \autoref{fig:APT_analysis_Ti3C2}~(a) shows the \ce{Ti3C2T_x} MXenes nanosheets with a complex 2D morphology, likely arising from agglomerated or folded nanosheets. These are embedded in \ce{Co}, shown in yellow. Differences in the evaporation field between the matrix and the nanosheets lead to an irregular curvature of the end surface of the specimen, which results in trajectory aberrations, varying magnification and intermixing zones in the reconstruction~\cite{1987Mil}. An indicative thickness at full width at half maximum of \SI{3.2}{\nano\metre} was obtained by evaluating the \ce{Co} 1D composition profile in \autoref{fig:APT_analysis_Ti3C2}~(b). The nominal thickness of a \ce{Ti3C2T_x} monolayer has been reported to be within the order of \SI{1}{\nano\metre} to \SI{1.5}{\nano\metre}~\cite{2015Wan, 2016Lip}, whereas the interlayer distance between two separate nanosheets can vary depending on the intercalated species~\cite{2018Lip}. A single nanosheet or stack of up to three \ce{Ti3C2T_x} nanosheets was hence analyzed.

A region of interest, delineated by the dark green-cyan iso-composition surface, encompassing regions in the 3D atom map containing over \SI{6}{\at} \ce{Ti}, in \autoref{fig:APT_analysis_Ti3C2}~(a), was extracted to exclude the \ce{Co} matrix and any \ce{CoO} layer from electrodeposition for compositional analysis. Compared to the nominal value of \num{1.5}, the atomic ratio of \ce{Ti} to \ce{C} was measured to be \num{2.2}. Losses of \ce{C} have been reported in the APT analysis of carbides~\cite{2011Thu, 2018Pen, 2023Ndi}, depending on the analysis conditions, and partly due to detector saturation associated with two \ce{^12C+} ions~\cite{2019Pen}. Importantly, it has been questioned whether most MAX phases and their derived MXenes are actually pure carbides or rather oxycarbides~\cite{2022Mic}. Pure carbides may be more resistant to oxidation than oxycarbides, as recent results suggest~\cite{2022Mic, 2021Mat}. 

The specimen was analyzed immediately after synthesis; however, a significant amount of \ce{O} was measured in the material, approximately \SI{46.7}{\at}. The presence of \ce{O} can be attributed to synthesis in an aqueous solution, as the highly reactive surface is immediately bonded to, for example, \ce{O} or \ce{OH}. Previous reports have shown that under certain controlled high vacuum conditions in the transmission electron microscope the \ce{O} surface terminations can reach a supersaturation level of \SI{41}{\at} while retaining the \ce{Ti3C2T_x} MXene nanosheet structure~\cite{2020Per}, and the MXenes directly used from the solution were unlikely to be already oxidized MXenes. However, the measured amount of \ce{O} was not entirely attributable to \ce{O} or \ce{OH} surface terminations, but was also associated with weakly bonded water molecules to these surface terminations. Besides ion intercalation, intercalation of water molecules is a crucial factor for the aqueous MXene delamination~\cite{2017Alh}. Intrinsically hygroscopic cations, such as \ce{Li+} and \ce{Na+}, which were also detected as discussed below, also increase the amount of intercalated water molecules~\cite{2018Shp}. Characteristic peaks in the mass spectrum at \num{17}, \num{18}, and \SI{19}{\dalton} confirmed the evaporation of \ce{H_{1-3}O+} molecular ions, while peaks at \num{81}, \num{82}, \num{83}, \num{84}, and \SI{85}{\dalton}, partially overlapping with peaks from \ce{TiO2^+}, were attributed to the evaporation of \ce{TiO(OH3)+}. Quantification of the \ce{H} content was deliberately omitted, as it is known to be extremely challenging from APT to distinguish between \ce{H} from the analysis chamber and the sample in form of \ce{OH} surface terminations or intercalated water molecules itself~\cite{2022Che}. In the future, isotopic labeling using heavy water (\ce{D2O}) during MXene synthesis may clarify the controversial question of the intrinsic existence of \ce{OH} terminations on the MXene surface~\cite{2019Per, 2021Nat, 2022Näs}.

Not entirely unexpected, several functional surface terminations, intercalated elements, and impurities including \ce{Cl}, \ce{F}, \ce{Li}, \ce{Na}, and \ce{Al} were also incorporated and could be quantified within the extracted region of interest of the \ce{Ti3C2T_x} MXenes. The incorporation of \ce{Cl} and \ce{F} results from the wet chemical etching process of \ce{Al} from the precursor MAX phase and the synthesis of delaminated MXenes. High concentrations of \ce{HCl} and \ce{LiF} are used in the wet chemical synthesis~\cite{2014Ghi, 2021Kim}. Here, the \ce{Cl} content was measured to be \SI{2.49}{\at}. Accurate quantification of the amount of \ce{F} was complicated by the overlapping peaks of \ce{F+} and \ce{H3O+} in the mass spectrum region at \SI{19}{\dalton}. However, \ce{TiF^2+} molecular ions were measured (see the mass spectrum in the Supporting Information), so that the \ce{F} content could be quantified to at least \SI{1.16}{\at}. Tuning the termination groups on the surface of the MXene may influence its oxidation stability, for instance, \ce{Cl}-terminated MXenes have been shown to have higher stability than \ce{F}-terminated MXenes~\cite{2019Lu}.

Despite vigorous washing protocols, residuals of \ce{Li} and \ce{Na} elements were still present in the analyzed material at \SI{0.02}{\at} and \SI{0.11}{\at}, respectively. The amount of \ce{Na} was obtained by means of a peak decomposition with the overlapping peak of \ce{^{46}Ti^2+} at \SI{23}{\dalton}. \ce{Na} likely originated as an impurity in the synthesis chemicals, as was similarly observed for \ce{MoS2} synthesized by wet chemistry~\cite{2020Kim}. For example, \ce{Na} is a known impurity even in high-purity \ce{Li} salts for battery applications~\cite{2021Fu, 2021Mer}, and here \ce{LiF} (\SI{99}{\percent}) was used during synthesis. It should be noted that the spatial resolution of APT~\cite{2022Gau} does not allow to conclude whether these alkali impurities are on the surface or integrated into the structure of a single layer MXene. Due to electrostatic effects, these elements could be absorbed on the surface to stabilize it~\cite{2021Doo}, or they could spontaneously intercalate between two MXene layers~\cite{2013Luk}, as is even desired in the case of \ce{Li} in the synthesis route used here. The presence of \ce{Li} or \ce{Na} atoms on the surface may also directly impact the properties by intercalation doping~\cite{2016Wan}, which should grant further targeted investigations.

The complete removal of residual \ce{Al} atoms is possible by harsh \ce{HF} etching, often confirmed by X-ray (photoelectron) spectroscopy~\cite{2022Sai}. Although the wet etching environment removes the \ce{Al} layers within the MAX phase by mixing \ce{HCl} and \ce{LiF} to form in situ \ce{HF}, and the desired MXenes were rinsed seven times with deionized water, a trace amount of \ce{Al} residual atoms at \SI{0.17}{\at} was still measured within the \ce{Ti3C2T_x} MXenes. 

\begin{figure}[!htb]
    \centering
    \includegraphics[width = 7.01 in]{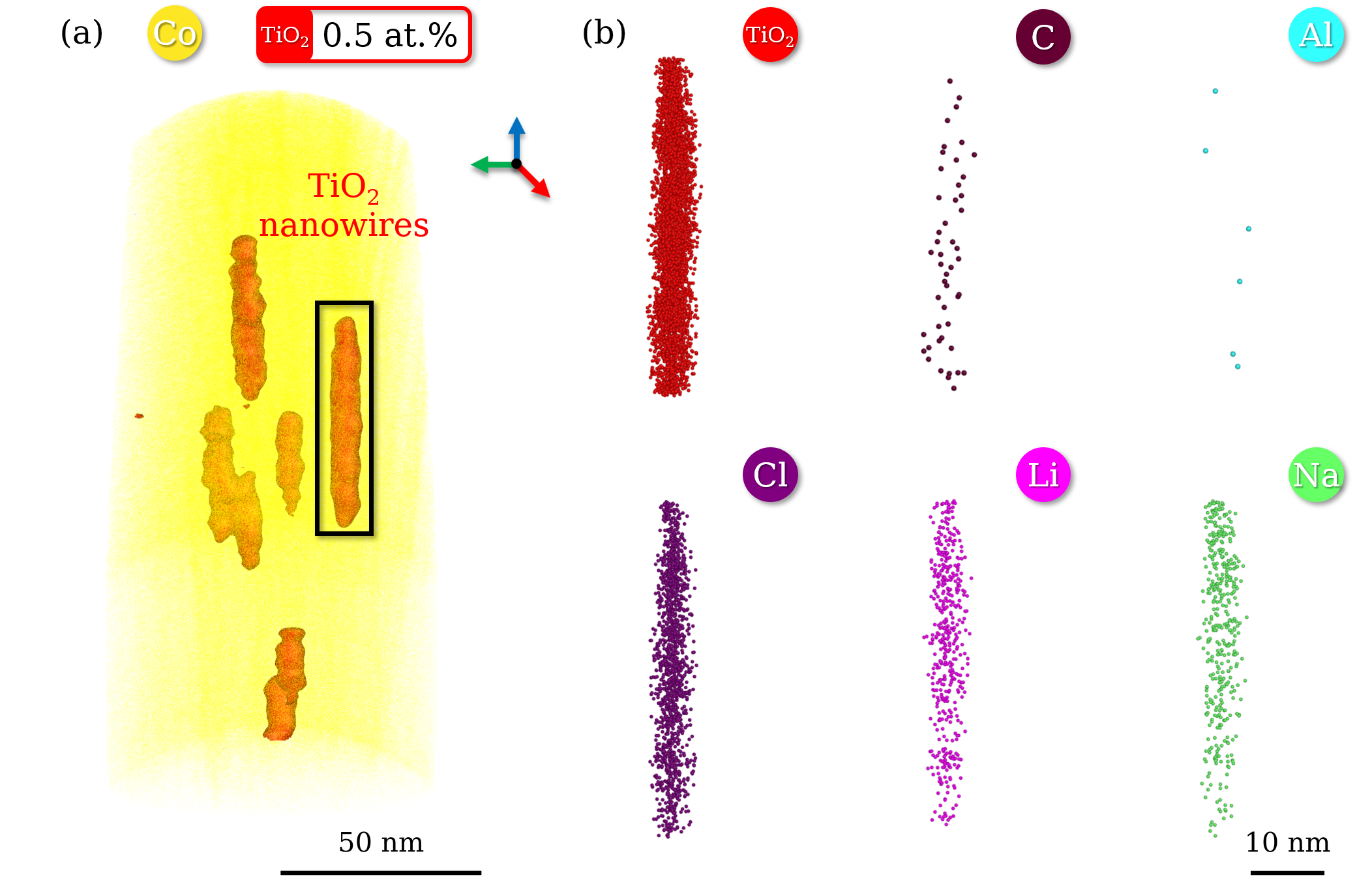}
    \caption{APT analysis of oxidized \ce{Ti3C2T_x} MXenes. (a) Reconstructed 3D atom map. \ce{TiO2} nanowires are highlighted by red iso-compositional surfaces at \SI{0.5}{\at}. (b) Molecular distribution map
    of \ce{TiO2} and elemental distribution maps of impurity elements in the extracted region of interest indicated in (a).}
    \label{fig:APT_analysis_TiO2}
\end{figure}

\autoref{fig:APT_analysis_TiO2}~(a) shows the reconstructed 3D atom map of the MXene oxidized in the colloidal solution. The structures highlighted by red iso-compositional surfaces of \SI{0.5}{\at} \ce{TiO2} are similar to those imaged by scanning transmission electron microscopy in \autoref{fig:Synthesis_Oxidation}~(c), which have a nanowire-like morphology. The atomic ratio of \ce{Ti} to \ce{O} within the extracted oxide particle, indicated by the black box, was measured to be \num{0.54}, which is close to the stoichiometric ratio of \ce{TiO2}. Changes in the acquisition parameters in APT~\cite{2019Ver} or a slight \ce{O} deficiency could be responsible for the off-stoichiometry of \ce{TiO2}. \autoref{fig:APT_analysis_TiO2}~(b) is a close-up of one of these nanowire-like structures with distribution maps of the individually detected elements. The measured impurity contents of \ce{C}, \ce{Cl}, \ce{F}, \ce{Li}, and \ce{Na} were \num{2.09}, \num{2.58}, \num{1.62}, \num{0.79}, and \SI{0.71}{\at}, respectively. \num{54 +- 30} atomic parts per million of \ce{Al} was collected in the oxidized MXene. 

\begin{figure}[!htb]
    \centering
    \includegraphics[width = 3.35 in]{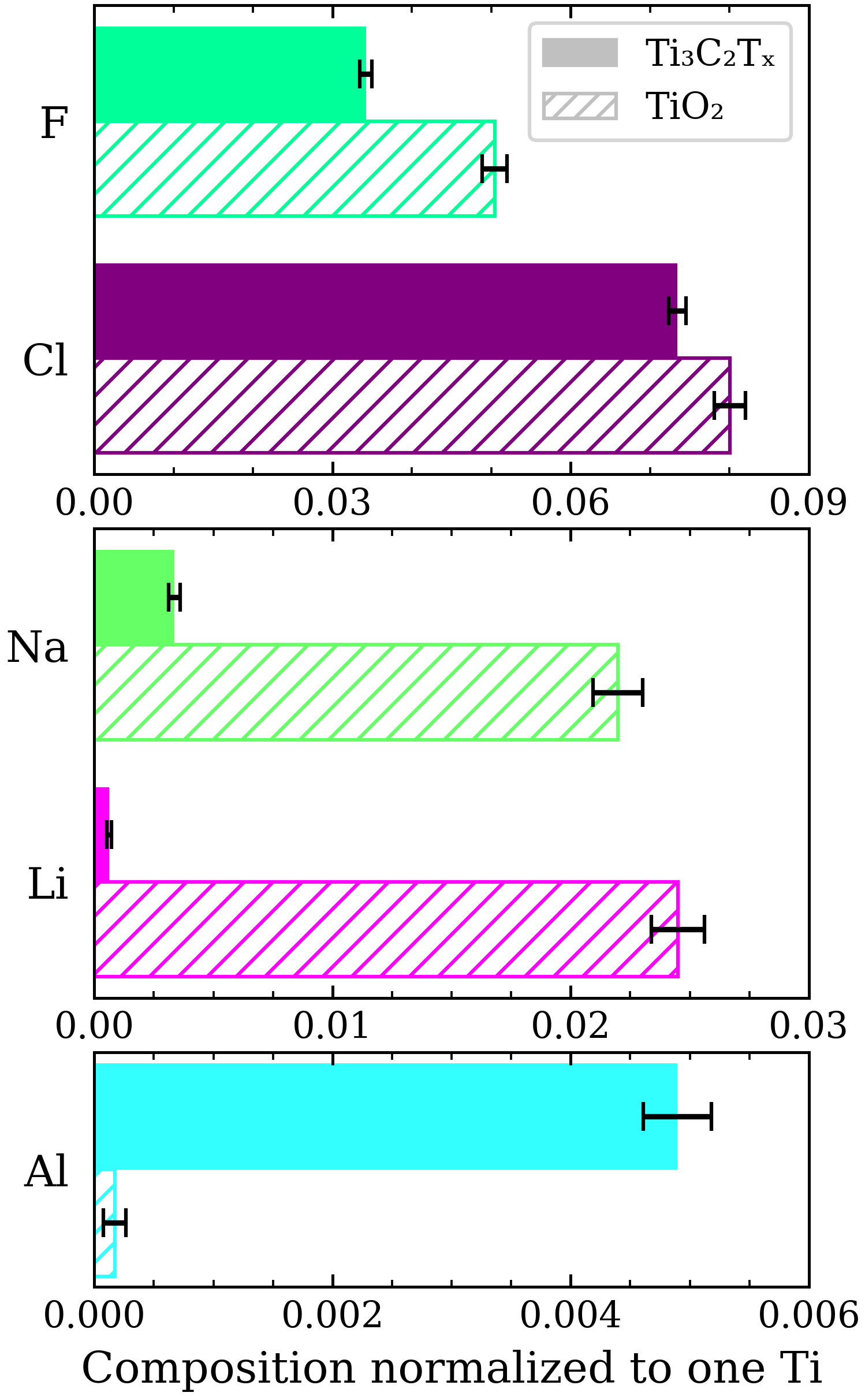}
    \caption{Comparison of the composition of the halogens, alkalies, and \ce{Al} in the as-synthesized \ce{Ti3C2T_x} MXenes and the oxidized \ce{TiO2}. For easier comparability of the data, the compositions were normalized to one \ce{Ti} (data given in \si{\at} are provided in the Supporting Information).}
    \label{fig:Summary_Impurities}
\end{figure}

Overall, the oxidation reaction of the MXene led to an enrichment of alkali elements and a depletion of unetched \ce{Al}, as visualized in \autoref{fig:Summary_Impurities}. Oxidation of \ce{Al} to \ce{Al2O3}~\cite{2014Xie} may be responsible for the depletion of \ce{Al}, as the \ce{Al2O3} and \ce{TiO2} phases have a positive mixing enthalpy~\cite{2017Zhe}, making the incorporation of \ce{Al} in the \ce{TiO2} phase unfavorable~\cite{2008Sun}. However, fundamental atomistic calculations comparing the stability of the incorporated alkali elements in the MXene and the oxide are missing. While the incorporation of \ce{Li} into anatase \ce{TiO2} is energetically feasible in general~\cite{2005Tie}, \ce{Na} is a known impurity of \ce{TiO2} synthesized on a soda-lime glass substrate~\cite{2016Xie}. Since several \ce{Li}~\cite{2018Wan} and \ce{Na} titanate derivatives~\cite{2017Dong, 2018Hua_NE, 2018Zhe, 2019Zho} were synthesized by oxidative treatment of \ce{Ti3C2T_x} MXenes with these alkali elements, it is reasonable to assume that they promote the formation of the oxide and stabilize it by lowering the free energy.   

Because metal-semiconductor heterostructures can provide rapid separation of photo-generated charge carriers~\cite{2021Yua}, \ce{Ti3C2T_x} MXenes were intentionally oxidized to form \ce{Ti3C2T_x}/\ce{TiO2} derivatives for photocatalysis~\cite{2018Low, 2020Che, 2021Deb, 2023Vid}. Other promising applications for these derivatives include electrodes for supercapacitors~\cite{2017Cao} and lithium-ion batteries~\cite{2023Jia}. However, in our previous report on \ce{TiO2} hollow nanowires, it was pointed out that the lack of precise characterization of impurities has an impact on the laboratory-synthesized \ce{TiO2} properties, which vary and are inconsistent with reported results~\cite{2020Lim}. For example, incorporated \ce{Na} influences the growth and crystallization of \ce{TiO2}, and thus the resulting properties~\cite{2016Xie}. From this point of view together with the observations made in this study, the alkali elements may have a significant influence on the oxidation mechanism and its kinetics, and consequently also on the properties of the \ce{Ti3C2T_x}/\ce{TiO2} derivatives.

In addition, the detected elements may also work as a dopant for both the as-synthesized \ce{Ti3C2T_x} MXenes and oxidized \ce{TiO2}. Doping is a proven and effective engineering strategy to enhance the performance of \ce{TiO2}, particularly with alkalies and halogens. For instance, \ce{Li}-doped \ce{TiO2} has shown significantly improved properties by increasing electronic conductivity, and thus faster electron transport~\cite{2020Tei}. Density functional theory calculations predicted an influence of \ce{Cl}- and \ce{F}-doping on the band gap of \ce{TiO2}~\cite{2019Fil} and incorporated \ce{Cl} shifts the absorption edge to a higher wavelength~\cite{2012Wan}, both effects enabling excellent photocatalytic activity. Doping \ce{TiO2} with \ce{C} can effectively facilitate photo-generated charge transfer and prevent electron-hole recombination~\cite{2020Han}. Surface terminations such as halogens are likely to influence the band structure, as functionalization of \ce{Ti3C2T_x} MXenes with halogens could lead to the formation of Dirac cones near the Fermi level, resulting in semi-metallic behavior~\cite{2021Far}. Experimentally, it was observed that preintercalated \ce{Na} leads to a reduced diffusion barrier for \ce{Na} ions and an increased number of active sites due to the increased interlayer spacing of the MXenes~\cite{2018Luo}. In summary, both alkali and halogen elements can affect the functional properties of \ce{Ti3C2T_x} MXenes and the oxidized \ce{TiO2}. It is therefore essential to study these incorporated elements on the near-atomic scale, for example using APT.

It has been recognized that the methods used to synthesize high-quality MXenes with chemical and structural stability can be critical. We have demonstrated the capability of APT to study the incorporation of alkalies and halogens as surface terminations, intercalated ions, or impurities introduced by the wet chemical synthesis of the MXenes. The alkali elements tended to further concentrate during the oxidation of the \ce{Ti3C2T_x} MXenes to \ce{TiO2} nanowires. Their presence may affect the physical properties of the MXenes themselves, but, importantly, this suggests that the presence and concentration of these elements may contribute to the oxidation mechanism and kinetics, by stabilizing the oxide to the detriment of the MXene itself. In the future, a broader understanding of the influence of these elements may enable targeted utilization to tailor the functional properties of the MXenes and their derived transition metal oxides.

\FloatBarrier


\section*{Supporting Information Available}

Synthesis of \ce{Ti3AlC2} MAX phase; Synthesis of \ce{Ti3C2T_x} MXenes; Electron Microscopy; APT specimen preparation; APT analysis; Mass spectrum analysis


\section*{Acknowledgment}

The authors acknowledge financial support from the German Research Foundation (DFG) through DIP Project No. 450800666. The support to the FIB and APT facilities at MPIE by Uwe Tezins, Andreas Sturm and Christian Broß is gratefully acknowledged.


\section*{Conflict of Interest}

The authors declare no conflict of interest.


\section*{Data Availability Statement}

The data that support the findings of this study are available from the corresponding author upon reasonable request.


\section*{Table of Contents}

Following oxidation of \ce{Ti3C2T_x} MXenes, atom probe tomography reveals that alkalies concentrate in \ce{TiO2} nanowires, indicating that their presence influences the oxidation mechanisms and kinetics. These results highlight how atom probe tomography can be utilized to better understand the functional surface terminations, intercalated ions or impurities that are inevitable in the wet chemical synthesis of MXenes.   

\begin{figure}[!htb]
    \centering
    \includegraphics[width=5.5cm, height=5cm, keepaspectratio]{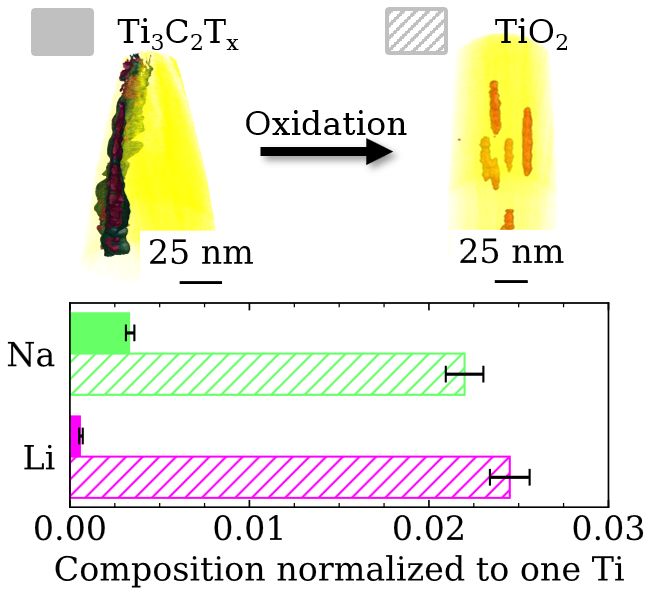}
\end{figure}


\printbibliography

@article{2011Nag,
author = {Naguib, Michael and Kurtoglu, Murat and Presser, Volker and Lu, Jun and Niu, Junjie and Heon, Min and Hultman, Lars and Gogotsi, Yury and Barsoum, Michel W.},
title = {Two-Dimensional Nanocrystals Produced by Exfoliation of Ti3AlC2},
journal = {Advanced Materials},
volume = {23},
number = {37},
pages = {4248-4253},
doi = {10.1002/adma.201102306},
year = {2011}
}

@Article{2017Ana,
author={Anasori, Babak and Lukatskaya, Maria R. and Gogotsi, Yury},
title={2D metal carbides and nitrides (MXenes) for energy storage},
journal={Nature Reviews Materials},
year={2017},
volume={2},
number={2},
pages={16098},
doi={10.1038/natrevmats.2016.98},
}

@Article{2022Li,
author={Li, Xinliang and Huang, Zhaodong and Shuck, Christopher E. and Liang, Guojin and Gogotsi, Yury and Zhi, Chunyi},
title={MXene chemistry, electrochemistry and energy storage applications},
journal={Nature Reviews Chemistry},
year={2022},
volume={6},
number={6},
pages={389-404},
doi={10.1038/s41570-022-00384-8},
}

@article{2021Mal,
author = {Maleski, Kathleen and Shuck, Christopher E. and Fafarman, Aaron T. and Gogotsi, Yury},
title = {The Broad Chromatic Range of Two-Dimensional Transition Metal Carbides},
journal = {Advanced Optical Materials},
volume = {9},
number = {4},
pages = {2001563},
doi = {10.1002/adom.202001563},
year = {2021}
}

@article{2019LiZhe,
author = {Li, Zhe and Wu, Yue},
title = {2D Early Transition Metal Carbides (MXenes) for Catalysis},
journal = {Small},
volume = {15},
number = {29},
pages = {1804736},
doi = {10.1002/smll.201804736},
year = {2019}
}

@Article{2018Hua_CSR,
author ="Huang, Kai and Li, Zhongjun and Lin, Jing and Han, Gang and Huang, Peng",
title  ="Two-dimensional transition metal carbides and nitrides (MXenes) for biomedical applications",
journal  ="Chemical Society Reviews",
year  ="2018",
volume  ="47",
issue  ="14",
pages  ="5109-5124",
doi  ="10.1039/C7CS00838D",
}

@article{2019Sol,
author = {Soleymaniha, Mohammadreza and Shahbazi, Mohammad-Ali and Rafieerad, Ali Reza and Maleki, Aziz and Amiri, Ahmad},
title = {Promoting Role of MXene Nanosheets in Biomedical Sciences: Therapeutic and Biosensing Innovations},
journal = {Advanced Healthcare Materials},
volume = {8},
number = {1},
pages = {1801137},
doi = {10.1002/adhm.201801137},
year = {2019}
}

@Article{2021Iqb,
author={Iqbal, Aamir and Hong, Junpyo and Ko, Tae Yun and Koo, Chong Min},
title={Improving oxidation stability of 2D MXenes: synthesis, storage media, and conditions},
journal={Nano Convergence},
year={2021},
volume={8},
number={1},
pages={9},
doi={10.1186/s40580-021-00259-6},
}

@Article{2022Jia,
author={Jiang, Jizhou and Bai, Saishuai and Zou, Jing and Liu, Song and Hsu, Jyh-Ping and Li, Neng and Zhu, Guoyin and Zhuang, Zechao and Kang, Qi and Zhang, Yizhou},
title={Improving stability of MXenes},
journal={Nano Research},
year={2022},
volume={15},
number={7},
pages={6551-6567},
doi={10.1007/s12274-022-4312-8},
}

@article{2022Cao,
author = {Cao, Fangcheng and Zhang, Ye and Wang, Hongqing and Khan, Karim and Tareen, Ayesha Khan and Qian, Wenjing and Zhang, Han and Ågren, Hans},
title = {Recent Advances in Oxidation Stable Chemistry of 2D MXenes},
journal = {Advanced Materials},
volume = {34},
number = {13},
pages = {2107554},
doi = {10.1002/adma.202107554},
year = {2022}
}

@ARTICLE{2019LiXin,
AUTHOR={Li, Xinliang and Huang, Zhaodong and Zhi, Chunyi},   
TITLE={Environmental Stability of MXenes as Energy Storage Materials},      
JOURNAL={Frontiers in Materials},      
VOLUME={6},           
YEAR={2019},      
DOI={10.3389/fmats.2019.00312},      
}

@article{2020Lia,
author = {Liang, Hongxing and Liu, Jing},
title = {Insights on the Corrosion and Degradation of MXenes as Electrocatalysts for Hydrogen Evolution Reaction},
journal = {ChemCatChem},
volume = {14},
number = {6},
pages = {e202101375},
doi = {10.1002/cctc.202101375},
year = {2022}
}

@Article{2021Bha,
author={Bhat, Anha and Anwer, Shoaib and Bhat, Kiesar Sideeq and Mohideen, M. Infas H. and Liao, Kin and Qurashi, Ahsanulhaq},
title={Prospects challenges and stability of 2D MXenes for clean energy conversion and storage applications},
journal={npj 2D Materials and Applications},
year={2021},
volume={5},
number={1},
pages={61},
doi={10.1038/s41699-021-00239-8},
}

@Article{2022Lim,
author={Lim, Kang Rui Garrick and Shekhirev, Mikhail and Wyatt, Brian C. and Anasori, Babak and Gogotsi, Yury and Seh, Zhi Wei},
title={Fundamentals of MXene synthesis},
journal={Nature Synthesis},
year={2022},
volume={1},
number={8},
pages={601-614},
doi={10.1038/s44160-022-00104-6},
}

@article{2021Nat,
title = {A critical analysis of the X-ray photoelectron spectra of Ti3C2Tz MXenes},
journal = {Matter},
volume = {4},
number = {4},
pages = {1224-1251},
year = {2021},
doi = {10.1016/j.matt.2021.01.015},
author = {Varun Natu and Mohamed Benchakar and Christine Canaff and Aurélien Habrioux and Stéphane Célérier and Michel W. Barsoum},
}

@article{2021Aln,
title = {Exploring MXenes and their MAX phase precursors by electron microscopy},
journal = {Materials Today Advances},
volume = {9},
pages = {100123},
year = {2021},
doi = {10.1016/j.mtadv.2020.100123},
author = {H. Alnoor and A. Elsukova and J. Palisaitis and I. Persson and E.N. Tseng and J. Lu and L. Hultman and P.O.{\AA}. Persson},
}

@article{2019Nat,
author = {Natu, Varun and Hart, James L. and Sokol, Maxim and Chiang, Helen and Taheri, Mitra L. and Barsoum, Michel W.},
title = {Edge Capping of 2D-MXene Sheets with Polyanionic Salts To Mitigate Oxidation in Aqueous Colloidal Suspensions},
journal = {Angewandte Chemie International Edition},
volume = {58},
number = {36},
pages = {12655-12660},
doi = {10.1002/anie.201906138},
year = {2019}
}

@Article{2020Lee,
author ="Lee, Yonghee and Kim, Seon Joon and Kim, Yong-Jae and Lim, Younghwan and Chae, Yoonjeong and Lee, Byeong-Joo and Kim, Young-Tae and Han, Hee and Gogotsi, Yury and Ahn, Chi Won",
title  ="Oxidation-resistant titanium carbide MXene films",
journal  ="Journal of Materials Chemistry A",
year  ="2020",
volume  ="8",
pages  ="573-581",
doi  ="10.1039/C9TA07036B",
}

@article{2022Ngu,
author = {Nguyen, Phuong Huyen and Nguyen, Duc Hieu and Kim, Donghyoung and Kim, Mun Kyoung and Jang, Jiseong and Sim, Woo Hyeong and Jeong, Hyung Mo and Namkoong, Gon and Jeong, Mun Seok},
title = {Regenerating MXene by a Facile Chemical Treatment Method},
journal = {ACS Applied Materials \& Interfaces},
volume = {14},
number = {45},
pages = {51487-51495},
year = {2022},
doi = {10.1021/acsami.2c13993},
}

@Article{2021Gau,
author={Gault, Baptiste and Chiaramonti, Ann and Cojocaru-Mir{\'e}din, Oana and Stender, Patrick and Dubosq, Renelle and Freysoldt, Christoph and Makineni, Surendra Kumar and Li, Tong and Moody, Michael and Cairney, Julie M.},
title={Atom probe tomography},
journal={Nature Reviews Methods Primers},
year={2021},
volume={1},
number={1},
pages={51},
doi={10.1038/s43586-021-00047-w},
}

@Article{2011Ted,
author={Tedsree, Karaked and Li, Tong and Jones, Simon and Chan, Chun Wong Aaron and Yu, Kai Man Kerry and Bagot, Paul A. J. and Marquis, Emmanuelle A. and Smith, George D. W. and Tsang, Shik Chi Edman},
title={Hydrogen production from formic acid decomposition at room temperature using a Ag--Pd core--shell nanocatalyst},
journal={Nature Nanotechnology},
year={2011},
volume={6},
number={5},
pages={302-307},
doi={10.1038/nnano.2011.42},
}

@article{2014Li,
author = {Li, Tong and Bagot, Paul A. J. and Christian, Elvis and Theobald, Brian R. C. and Sharman, Jonathan D. B. and Ozkaya, Dogan and Moody, Michael P. and Tsang, S. C. Edman and Smith, George D. W.},
title = {Atomic Imaging of Carbon-Supported Pt, Pt/Co, and Ir@Pt Nanocatalysts by Atom-Probe Tomography},
journal = {ACS Catalysis},
volume = {4},
number = {2},
pages = {695-702},
year = {2014},
doi = {10.1021/cs401117e},
}

@article{2014Fel,
author = {Felfer, Peter and Benndorf, Paul and Masters, Anthony and Maschmeyer, Thomas and Cairney, Julie M.},
title = {Revealing the Distribution of the Atoms within Individual Bimetallic Catalyst Nanoparticles},
journal = {Angewandte Chemie International Edition},
volume = {53},
number = {42},
pages = {11190-11193},
doi = {10.1002/anie.201405043},
year = {2014}
}

@article{2006Per,
author = {Perea, Daniel E. and Allen, Jonathan E. and May, Steven J. and Wessels, Bruce W. and Seidman, David N. and Lauhon, Lincoln J.},
title = {Three-Dimensional Nanoscale Composition Mapping of Semiconductor Nanowires},
journal = {Nano Letters},
volume = {6},
number = {2},
pages = {181-185},
year = {2006},
doi = {10.1021/nl051602p},
}

@article{2020Lim,
author = {Lim, Joohyun and Kim, Se-Ho and Aymerich Armengol, Raquel and Kasian, Olga and Choi, Pyuck-Pa and Stephenson, Leigh T. and Gault, Baptiste and Scheu, Christina},
title = {Atomic-Scale Mapping of Impurities in Partially Reduced Hollow TiO2 Nanowires},
journal = {Angewandte Chemie International Edition},
volume = {59},
number = {14},
pages = {5651-5655},
doi = {10.1002/anie.201915709},
year = {2020}
}

@article{2020Kim,
author = {Kim, Se-Ho and Lim, Joohyun and Sahu, Rajib and Kasian, Olga and Stephenson, Leigh T. and Scheu, Christina and Gault, Baptiste},
title = {Direct Imaging of Dopant and Impurity Distributions in 2D MoS2},
journal = {Advanced Materials},
volume = {32},
number = {8},
pages = {1907235},
doi = {10.1002/adma.201907235},
year = {2020}
}

@article{2022Kim_JACS,
author = {Kim, Se-Ho and Yoo, Su-Hyun and Chakraborty, Poulami and Jeong, Jiwon and Lim, Joohyun and El-Zoka, Ayman A. and Zhou, Xuyang and Stephenson, Leigh T. and Hickel, Tilmann and Neugebauer, Jörg and Scheu, Christina and Todorova, Mira and Gault, Baptiste},
title = {Understanding Alkali Contamination in Colloidal Nanomaterials to Unlock Grain Boundary Impurity Engineering},
journal = {Journal of the American Chemical Society},
volume = {144},
number = {2},
pages = {987-994},
year = {2022},
doi = {10.1021/jacs.1c11680},
}

@article{2022Kim_AM,
author = {Kim, Se-Ho and Yoo, Su-Hyun and Shin, Sangyong and El-Zoka, Ayman A. and Kasian, Olga and Lim, Joohyun and Jeong, Jiwon and Scheu, Christina and Neugebauer, Jörg and Lee, Hyunjoo and Todorova, Mira and Gault, Baptiste},
title = {Controlled Doping of Electrocatalysts through Engineering Impurities},
journal = {Advanced Materials},
volume = {34},
number = {28},
pages = {2203030},
doi = {10.1002/adma.202203030},
year = {2022}
}

@article{2018Kim,
title = {A new method for mapping the three-dimensional atomic distribution within nanoparticles by atom probe tomography (APT)},
journal = {Ultramicroscopy},
volume = {190},
pages = {30-38},
year = {2018},
doi = {10.1016/j.ultramic.2018.04.005},
author = {Se-Ho Kim and Phil Woong Kang and O Ok Park and Jae-Bok Seol and Jae-Pyoung Ahn and Ji Yeong Lee and Pyuck-Pa Choi},
}

@article{2021Wes,
author = {Olivia Westhead  and Rhodri Jervis  and Ifan E. L. Stephens },
title = {Is lithium the key for nitrogen electroreduction?},
journal = {Science},
volume = {372},
number = {6547},
pages = {1149-1150},
year = {2021},
doi = {10.1126/science.abi8329},
}

@Article{2014Gha,
author ="Ghassemi, H. and Harlow, W. and Mashtalir, O. and Beidaghi, M. and Lukatskaya, M. R. and Gogotsi, Y. and Taheri, M. L.",
title  ="In situ environmental transmission electron microscopy study of oxidation of two-dimensional Ti3C2 and formation of carbon-supported TiO2",
journal  ="Journal of Materials Chemistry A",
year  ="2014",
volume  ="2",
issue  ="35",
pages  ="14339-14343",
doi  ="10.1039/C4TA02583K",
}

@Article{2019Fan,
author ="Xia, Fanjie and Lao, Junchao and Yu, Ruohan and Sang, Xiahan and Luo, Jiayan and Li, Yu and Wu, Jinsong",
title  ="Ambient oxidation of Ti3C2 MXene initialized by atomic defects",
journal  ="Nanoscale",
year  ="2019",
volume  ="11",
issue  ="48",
pages  ="23330-23337",
doi  ="10.1039/C9NR07236E",
}

@Article{2019Cha,
author ="Chae, Yoonjeong and Kim, Seon Joon and Cho, Soo-Yeon and Choi, Junghoon and Maleski, Kathleen and Lee, Byeong-Joo and Jung, Hee-Tae and Gogotsi, Yury and Lee, Yonghee and Ahn, Chi Won",
title  ="An investigation into the factors governing the oxidation of two-dimensional Ti3C2 MXene",
journal  ="Nanoscale",
year  ="2019",
volume  ="11",
issue  ="17",
pages  ="8387-8393",
doi  ="10.1039/C9NR00084D",
}

@article{2017Zha,
author = {Zhang, Chuanfang John and Pinilla, Sergio and McEvoy, Niall and Cullen, Conor P. and Anasori, Babak and Long, Edmund and Park, Sang-Hoon and Seral-Ascaso, Andrés and Shmeliov, Aleksey and Krishnan, Dileep and Morant, Carmen and Liu, Xinhua and Duesberg, Georg S. and Gogotsi, Yury and Nicolosi, Valeria},
title = {Oxidation Stability of Colloidal Two-Dimensional Titanium Carbides (MXenes)},
journal = {Chemistry of Materials},
volume = {29},
number = {11},
pages = {4848-4856},
year = {2017},
doi = {10.1021/acs.chemmater.7b00745},
}

@article{2021Jun,
title={Atom Probe Tomography Investigations of Ag Nanoparticles Embedded in Pulse-Electrodeposited Ni Films},
volume={27},
DOI={10.1017/S1431927621012137},
number={5},
journal={Microscopy and Microanalysis},
author={Jun, Hosun and Jang, Kyuseon and Jung, Chanwon and Choi, Pyuck-Pa},
year={2021},
pages={1007–1016},
}

@article{2011Thu,
title = {Quantitative atom probe analysis of carbides},
journal = {Ultramicroscopy},
volume = {111},
number = {6},
pages = {604-608},
year = {2011},
doi = {10.1016/j.ultramic.2010.12.024},
author = {M. Thuvander and J. Weidow and J. Angseryd and L.K.L. Falk and F. Liu and M. Sonestedt and K. Stiller and H.-O. Andrén},
}

@article{2018Pen,
title = {On the detection of multiple events in atom probe tomography},
journal = {Ultramicroscopy},
volume = {189},
pages = {54-60},
year = {2018},
doi = {10.1016/j.ultramic.2018.03.018},
author = {Zirong Peng and Francois Vurpillot and Pyuck-Pa Choi and Yujiao Li and Dierk Raabe and Baptiste Gault},
}

@article{2019Pen,
author = {Peng, Zirong and Zanuttini, David and Gervais, Benoit and Jacquet, Emmanuelle and Blum, Ivan and Choi, Pyuck-Pa and Raabe, Dierk and Vurpillot, Francois and Gault, Baptiste},
title = {Unraveling the Metastability of Cn2+ (n = 2–4) Clusters},
journal = {The Journal of Physical Chemistry Letters},
volume = {10},
number = {3},
pages = {581-588},
year = {2019},
doi = {10.1021/acs.jpclett.8b03449},
}

@article{2020Per,
author = {Persson, Ingemar and Halim, Joseph and Hansen, Thomas W. and Wagner, Jakob B. and Darakchieva, Vanya and Palisaitis, Justinas and Rosen, Johanna and Persson, Per O. {\AA}.},
title = {How Much Oxygen Can a MXene Surface Take Before It Breaks?},
journal = {Advanced Functional Materials},
volume = {30},
number = {47},
pages = {1909005},
doi = {10.1002/adfm.201909005},
year = {2020}
}

@Article{2014Ghi,
author={Ghidiu, Michael and Lukatskaya, Maria R. and Zhao, Meng-Qiang and Gogotsi, Yury and Barsoum, Michel W.},
title={Conductive two-dimensional titanium carbide `clay' with high volumetric capacitance},
journal={Nature},
year={2014},
volume={516},
number={7529},
pages={78-81},
doi={10.1038/nature13970},
}

@article{2021Kim,
author = {Kim, Yong-Jae and Kim, Seon Joon and Seo, Darae and Chae, Yoonjeong and Anayee, Mark and Lee, Yonghee and Gogotsi, Yury and Ahn, Chi Won and Jung, Hee-Tae},
title = {Etching Mechanism of Monoatomic Aluminum Layers during MXene Synthesis},
journal = {Chemistry of Materials},
volume = {33},
number = {16},
pages = {6346-6355},
year = {2021},
doi = {10.1021/acs.chemmater.1c01263},
}

@Article{2019Lu,
author ="Lu, J. and Persson, I. and Lind, H. and Palisaitis, J. and Li, M. and Li, Y. and Chen, K. and Zhou, J. and Du, S. and Chai, Z. and Huang, Z. and Hultman, L. and Eklund, P. and Rosen, J. and Huang, Q. and Persson, P. O.{\AA}.",
title  ="Tin+1Cn MXenes with fully saturated and thermally stable Cl terminations",
journal  ="Nanoscale Advances",
year  ="2019",
volume  ="1",
issue  ="9",
pages  ="3680-3685",
doi  ="10.1039/C9NA00324J",
}

@article{2022Gau,
    author = {Gault, Baptiste and Klaes, Benjamin and Morgado, Felipe F and Freysoldt, Christoph and Li, Yue and De Geuser, Frederic and Stephenson, Leigh T and Vurpillot, François},
    title = "{Reflections on the Spatial Performance of Atom Probe Tomography in the Analysis of Atomic Neighborhoods}",
    journal = {Microscopy and Microanalysis},
    volume = {28},
    number = {4},
    pages = {1116-1126},
    year = {2022},
    doi = {10.1017/S1431927621012952},
}

@article{2021Doo,
author = {Doo, Sehyun and Chae, Ari and Kim, Daesin and Oh, Taegon and Ko, Tae Yun and Kim, Seon Joon and Koh, Dong-Yeun and Koo, Chong Min},
title = {Mechanism and Kinetics of Oxidation Reaction of Aqueous Ti3C2Tx Suspensions at Different pHs and Temperatures},
journal = {ACS Applied Materials \& Interfaces},
volume = {13},
number = {19},
pages = {22855-22865},
year = {2021},
doi = {10.1021/acsami.1c04663},
}

@article{2013Luk,
author = {Maria R. Lukatskaya  and Olha Mashtalir  and Chang E. Ren  and Yohan Dall’Agnese  and Patrick Rozier  and Pierre Louis Taberna  and Michael Naguib  and Patrice Simon  and Michel W. Barsoum  and Yury Gogotsi },
title = {Cation Intercalation and High Volumetric Capacitance of Two-Dimensional Titanium Carbide},
journal = {Science},
volume = {341},
number = {6153},
pages = {1502-1505},
year = {2013},
doi = {10.1126/science.1241488},
}

@ARTICLE{2022Sai,
AUTHOR={Saita, Emi and Iwata, Masaki and Shibata, Yuki and Matsunaga, Yuki and Suizu, Rie and Awaga, Kunio and Hirotani, Jun and Omachi, Haruka},   
TITLE={Exfoliation of Al-Residual Multilayer MXene Using Tetramethylammonium Bases for Conductive Film Applications},      
JOURNAL={Frontiers in Chemistry},      
VOLUME={10},           
YEAR={2022},      
DOI={10.3389/fchem.2022.841313},      
}

@article{2019Ver,
    author = {Verberne, Rick and Saxey, David W and Reddy, Steven M and Rickard, William D A and Fougerouse, Denis and Clark, Chris},
    title = "{Analysis of Natural Rutile (TiO2) by Laser-assisted Atom Probe Tomography}",
    journal = {Microscopy and Microanalysis},
    volume = {25},
    number = {2},
    pages = {539-546},
    year = {2019},
    doi = {10.1017/S1431927618015477},
}

@article{2017Zhe,
  title={Solid-state Reaction Studies in Al2O3–TiO2 System by Diffusion Couple Method},
  author={Jianchao Zheng and Xiaojun Hu and Zhongshan Ren and Xiangxin Xue and Kuochih Chou},
  journal={ISIJ International},
  volume={57},
  number={10},
  pages={1762-1766},
  year={2017},
  doi={10.2355/isijinternational.ISIJINT-2017-042}
}

@Article{2008Sun,
author={Sun, Min-Ki and Jung, In-Ho and Lee, Hae-Geon},
title={Morphology and chemistry of oxide inclusions after Al and Ti complex deoxidation},
journal={Metals and Materials International},
year={2008},
volume={14},
number={6},
pages={791-798},
doi={10.3365/met.mat.2008.12.791},
}

@article{2021Yua,
author = {Yuan, Lan and Geng, Zhaoyi and Xu, Jikun and Guo, Fen and Han, Chuang},
title = {Metal-Semiconductor Heterostructures for Photoredox Catalysis: Where Are We Now and Where Do We Go?},
journal = {Advanced Functional Materials},
volume = {31},
number = {27},
pages = {2101103},
doi = {10.1002/adfm.202101103},
year = {2021}
}

@article{2018Low,
title = {TiO2/MXene Ti3C2 composite with excellent photocatalytic CO2 reduction activity},
journal = {Journal of Catalysis},
volume = {361},
pages = {255-266},
year = {2018},
doi = {10.1016/j.jcat.2018.03.009},
author = {Jingxiang Low and Liuyang Zhang and Tong Tong and Baojia Shen and Jiaguo Yu},
}

@article{2020Che,
title = {Morphology and photocatalytic activity of TiO2/MXene composites by in-situ solvothermal method},
journal = {Ceramics International},
volume = {46},
number = {12},
pages = {20088-20096},
year = {2020},
doi = {10.1016/j.ceramint.2020.05.083},
author = {Jin Chen and Huiqi Zheng and Yang Zhao and Meidan Que and Wendong Wang and Xiping Lei},
}

@article{2021Deb,
author = {Debow, Shaun and Zhang, Tong and Liu, Xusheng and Song, Fuzhan and Qian, Yuqin and Han, Jian and Maleski, Kathleen and Zander, Zachary B. and Creasy, William R. and Kuhn, Danielle L. and Gogotsi, Yury and DeLacy, Brendan G. and Rao, Yi},
title = {Charge Dynamics in TiO2/MXene Composites},
journal = {The Journal of Physical Chemistry C},
volume = {125},
number = {19},
pages = {10473-10482},
year = {2021},
doi = {10.1021/acs.jpcc.1c01543},
}

@Article{2023Vid,
author ="Vida, Július and Gemeiner, Pavol and Pavličková, Michaela and Mazalová, Martina and Souček, Pavel and Plašienka, Dušan and Homola, Tomáš",
title  ="Nanocrystalline TiO2/Ti3C2Tx MXene composites with a tunable work function prepared using atmospheric pressure oxygen plasma",
journal  ="Nanoscale",
year  ="2023",
volume  ="15",
issue  ="3",
pages  ="1289-1298",
doi  ="10.1039/D2NR04465J",
}

@article{2017Cao,
doi = {10.1149/2.1541714jes},
year = {2017},
volume = {164},
number = {14},
pages = {A3933},
author = {Minjuan Cao and Fen Wang and Lei Wang and Wenling Wu and Wenjing Lv and Jianfeng Zhu},
title = {Room Temperature Oxidation of Ti3C2 MXene for Supercapacitor Electrodes},
journal = {Journal of The Electrochemical Society},
}

@Article{2023Jia,
author={Jia, Yajuan and Liu, Junhui and Shang, Lisha},
title={Layered structure 2D MXene/TiO2 composites as high-performance anodes for lithium-ion batteries},
journal={Ionics},
year={2023},
volume={29},
number={2},
pages={531-537},
doi={10.1007/s11581-022-04862-3},
}

@article{2020Tei,
title = {Synthesizing Li doped TiO2 electron transport layers for highly efficient planar perovskite solar cell},
journal = {Superlattices and Microstructures},
volume = {145},
pages = {106627},
year = {2020},
doi = {10.1016/j.spmi.2020.106627},
author = {Razieh Teimouri and Zahra Heydari and Mohammad Pouya Ghaziani and Mahdi Madani and Hamed Abdy and Mohammadreza Kolahdouz and Ebrahim Asl-Soleimani},
}

@article{2020Han,
title = {Ti3C2 MXene-derived carbon-doped TiO2 coupled with g-C3N4 as the visible-light photocatalysts for photocatalytic H2 generation},
journal = {Applied Catalysis B: Environmental},
volume = {265},
pages = {118539},
year = {2020},
doi = {10.1016/j.apcatb.2019.118539},
author = {Xin Han and Lin An and Yue Hu and Yaogang Li and Chengyi Hou and Hongzhi Wang and Qinghong Zhang},
}

@article{2016Xie,
author = {Xie, Huan and Li, Neng and Liu, Baoshun and Yang, Jingjing and Zhao, Xiujian},
title = {Role of Sodium Ion on TiO2 Photocatalyst: Influencing Crystallographic Properties or Serving as the Recombination Center of Charge Carriers?},
journal = {The Journal of Physical Chemistry C},
volume = {120},
number = {19},
pages = {10390-10399},
year = {2016},
doi = {10.1021/acs.jpcc.6b01730},
}

@article{2012Wan,
title = {Sonochemical synthesis and characterization of Cl-doped TiO2 and its application in the photodegradation of phthalate ester under visible light irradiation},
journal = {Chemical Engineering Journal},
volume = {189-190},
pages = {288-294},
year = {2012},
doi = {10.1016/j.cej.2012.02.078},
author = {Xi-Kui Wang and Chen Wang and Wen-Qiang Jiang and Wei-Lin Guo and Jing-Gang Wang},
}

@Article{2019Fil,
author={Filippatos, Petros-Panagis and Kelaidis, Nikolaos and Vasilopoulou, Maria and Davazoglou, Dimitris and Lathiotakis, Nektarios N. and Chroneos, Alexander},
title={Defect processes in F and Cl doped anatase TiO2},
journal={Scientific Reports},
year={2019},
volume={9},
number={1},
pages={19970},
doi={10.1038/s41598-019-55518-8},
}

@Article{2016Wan,
author ="Wan, Jiayu and Lacey, Steven D. and Dai, Jiaqi and Bao, Wenzhong and Fuhrer, Michael S. and Hu, Liangbing",
title  ="Tuning two-dimensional nanomaterials by intercalation: materials{,} properties and applications",
journal  ="Chemical Society Reviews",
year  ="2016",
volume  ="45",
issue  ="24",
pages  ="6742-6765",
doi  ="10.1039/C5CS00758E",
}

@article{2023Ndi,
author = {Ndiaye, Samba and Bacchi, Christian and Klaes, Benjamin and Canino, Mariaconcetta and Vurpillot, François and Rigutti, Lorenzo},
title = {Surface Dynamics of Field Evaporation in Silicon Carbide},
journal = {The Journal of Physical Chemistry C},
volume = {127},
number = {11},
pages = {5467-5478},
year = {2023},
doi = {10.1021/acs.jpcc.2c08908},
}

@Article{2022Mic,
author={Micha{\l}owski, Pawe{\l} P. and Anayee, Mark and Mathis, Tyler S. and Kozdra, Sylwia and W{\'o}jcik, Adrianna and Hantanasirisakul, Kanit and J{\'o}{\'{z}}wik, Iwona and Pi{\k{a}}tkowska, Anna and Mo{\.{z}}d{\.{z}}onek, Ma{\l}gorzata and Malinowska, Agnieszka and Diduszko, Ryszard and Wierzbicka, Edyta and Gogotsi, Yury},
title={Oxycarbide MXenes and MAX phases identification using monoatomic layer-by-layer analysis with ultralow-energy secondary-ion mass spectrometry},
journal={Nature Nanotechnology},
year={2022},
volume={17},
number={11},
pages={1192-1197},
doi={10.1038/s41565-022-01214-0},
}

@article{2021Mat,
author = {Mathis, Tyler S. and Maleski, Kathleen and Goad, Adam and Sarycheva, Asia and Anayee, Mark and Foucher, Alexandre C. and Hantanasirisakul, Kanit and Shuck, Christopher E. and Stach, Eric A. and Gogotsi, Yury},
title = {Modified MAX Phase Synthesis for Environmentally Stable and Highly Conductive Ti3C2 MXene},
journal = {ACS Nano},
volume = {15},
number = {4},
pages = {6420-6429},
year = {2021},
doi = {10.1021/acsnano.0c08357},
}

@article{2017Alh,
author = {Alhabeb, Mohamed and Maleski, Kathleen and Anasori, Babak and Lelyukh, Pavel and Clark, Leah and Sin, Saleesha and Gogotsi, Yury},
title = {Guidelines for Synthesis and Processing of Two-Dimensional Titanium Carbide (Ti3C2Tx MXene)},
journal = {Chemistry of Materials},
volume = {29},
number = {18},
pages = {7633-7644},
year = {2017},
doi = {10.1021/acs.chemmater.7b02847},
}

@article{2018Shp,
author = {Shpigel, Netanel and Levi, Mikhael D. and Sigalov, Sergey and Mathis, Tyler S. and Gogotsi, Yury and Aurbach, Doron},
title = {Direct Assessment of Nanoconfined Water in 2D Ti3C2 Electrode Interspaces by a Surface Acoustic Technique},
journal = {Journal of the American Chemical Society},
volume = {140},
number = {28},
pages = {8910-8917},
year = {2018},
doi = {10.1021/jacs.8b04862},
}

@article{2022Näs,
title = {XPS spectra curve fittings of Ti3C2Tx based on first principles thinking},
journal = {Applied Surface Science},
volume = {593},
pages = {153442},
year = {2022},
issn = {0169-4332},
doi = {10.1016/j.apsusc.2022.153442},
author = {Lars-Åke Näslund and Ingemar Persson},
}

@article{1987Mil,
  title={The effects of local magnification and trajectory aberrations on atom probe analysis},
  author={Michael K. Miller},
  journal={Le Journal De Physique Colloques},
  year={1987},
  volume={48},
  pages={565-570}
}


\end{document}